\documentclass[preprint]{aastex}

\begin{document}
\title{A Subaru/HDS study of Lead (Pb) abundances in eight {\it
s}-process-element-rich, metal-poor stars\thanks{Based on data
collected at the Subaru Telescope, which is operated by the National
Astronomical Observatory of Japan.}}
\author{Wako Aoki\altaffilmark{1}, Sean G. Ryan\altaffilmark{2}, John
E. Norris\altaffilmark{3}, Timothy C. Beers\altaffilmark{4}, Hiroyasu
Ando\altaffilmark{1}, Stelios Tsangarides\altaffilmark{2}}
\altaffiltext{1}{National Astronomical Observatory, Mitaka, Tokyo,
181-8588 Japan; email: aoki.wako@nao.ac.jp, ando@optik.mtk.nao.ac.jp}
\altaffiltext{2}{Department of Physics and Astronomy, The Open
University, Walton Hall, Milton Keynes, MK7 6AA, UK; email:
s.g.ryan@open.ac.uk, s.tsangarides@open.ac.uk}
\altaffiltext{3}{Research School of Astronomy \&
Astrophysics, The Australian National University, Mount Stromlo Observatory,
Cotter Road, Weston, ACT 2611, Australia; jen@mso.anu.edu.au}
\altaffiltext{4}{Department of Physics and Astronomy, Michigan State
University, East Lansing, MI 48824-1116; email: beers@pa.msu.edu}

\begin{abstract} 

We report the abundances of neutron-capture elements in eight carbon-rich,
metal-poor ($-2.7\leq$[Fe/H]$\leq -1.9$) stars observed with the Subaru
Telescope High Dispersion Spectrograph. The derived abundance patterns indicate
that the neutron-capture elements in these objects primarily originated from
{\it s}-process nucleosynthesis, although the [Ba/Eu] abundance ratios in some
objects are lower than that of the solar-system {\it s}-process component. The
present analysis has yielded the Pb abundances for seven objects, as well as an
upper limit for one object, from use of the \ion{Pb}{1} 4057~{\AA} and
3683~{\AA} lines.  The values of [Pb/Ba] in these objects cover a wide range,
between $-0.3$ and +1.2. Theoretical studies of {\it s}-process nucleosynthesis
at low metallicity are required to explain this large dispersion of the [Pb/Ba]
values. Variations in radial velocity have been found for two of the eight
objects, suggesting that, at least in these instances, the observed excess of
{\it s}-process elements is due to the transfer of material across a binary
system including an AGB star.  Comparisons with predictions of AGB
nucleosynthesis models are discussed.

\end{abstract}


\section{Introduction}\label{sec:intro}

Lead (Pb) isotopes form, along with those of bismuth, the group of the heaviest
stable nuclei.  Understanding of the physical processes responsible for the
synthesis of these nuclei, as well as their enrichment history in the Galaxy,
is important in the study of stellar nucleosynthesis and Galactic chemical
evolution, as well as for precision cosmochronology, since these nuclei are the
expected decay products of Th and U \citep{goriely01, schatz02}.   
The majority
of Pb nuclei in the solar system are believed to have originated in the {\it
s}-process.  However, the so-called main-component of the {\it s}-process in
classical models cannot explain the solar system abundances of Pb, hence
another component (the so-called strong component) was introduced
\citep[e.g., ][]{kappeler89}. Recent theoretical work on nucleosynthesis in
asymptotic giant branch (AGB) stars, based on the idea of an {\it s}-process
that occurs in the radiative zone during the inter-pulse phase, have predicted
large production of Pb and Bi ($Z=82$ and 83, respectively) compared with
lighter {\it s}-process elements such as Sr ($Z=38$) and Ba ($Z=56$) in low
metallicity AGB stars \citep{gallino98, goriely00, busso01}. These works
suggested that the Pb production in low-metallicity AGB stars corresponds
to the strong component assumed in the classical models. Studies of Pb
enrichment in the Galaxy have also started using the yields predicted by these
AGB models \citep{travaglio01}.

Recent abundance studies of ${\it s}$-process-element-enhanced, metal-poor
stars, based on high-resolution spectroscopy, have measured the abundances of
Pb and other neutron-capture elements in a number of stars, and made it
possible to examine the models of AGB nucleosynthesis in greater detail.  For
instance, \citet{aoki00} determined the abundances of 16 neutron-capture
elements, including Pb, for the ${\it s}$-process-element-rich, very metal-poor
([Fe/H]=$-$2.7) star LP~625--44\footnote{[A/B] = $\log(N_{\rm A}/N_{\rm B})-
\log(N_{\rm A}/N_{\rm B})_{\odot}$, and $\log \epsilon_{\rm A} = \log(N_{\rm
A}/N_{\rm H})+12$ for elements A and B.}. The Pb abundance of this object is,
however, much {\it lower} than the prediction by the ``standard'' model of
\citet{gallino98}, when the abundance pattern is normalized by the abundance of
lighter {\it s}-process nuclei (e.g., Ba). A similar result was derived for
another metal-poor star, LP~706--7, by \citet{aoki01}. Their abundances could,
however, be reconciled with the models of Gallino et al. if the extent of the
$^{13}$C pocket is reduced by a factor of $\simeq$24 \citep{ryan01}.  On the
other hand, \citet{vaneck01} studied another three {\it
s}-process-element-enhanced, metal-poor objects, and found very {\it large} Pb
enhancements in these objects, compared to those of the lighter elements.  They
concluded that the results can be explained well by the model of AGB
nucleosynthesis presented by \citet{goriely00}.

Thus, the abundance ratios of the Pb, and the neutron-capture elements
at the second peak of the {\it s}-process ($Z\sim 56$), produced by
metal-deficient AGB stars seem to show a large dispersion, or might
even be classified into two distinct groups. In order to understand
the nature of the {\it s}-process nucleosynthesis at low metallicity,
we have embarked on an extensive set of studies of the neutron-capture
elements for metal-poor stars with excesses of {\it s}-process
elements. In this Paper we report the abundances of neutron-capture
elements, including Pb, for eight metal-poor stars.

\section{Observations and Measurements}\label{sec:obs}

We selected our program sample from the list of metal-poor stars provided by
the HK survey of Beers and collaborators \citep{beers92, beers99, norris99}, taking
account of the metallicity and apparent level of carbon enhancement (as
indicated by an index, $GP$, that measures the strength of the CH G band at
4300 \AA\ ).  We have observed more than 10 stars, and selected the 7 stars
that clearly exhibit excesses of the {\it s}-process elements.  For purposes of
comparison, we have also observed the {\it s}-process-element-rich star
HD~196944 studied by \citet{vaneck01}.

Observations were carried out with the High Dispersion Spectrograph
(HDS) of the 8.2~m Subaru Telescope \citep{noguchi02}. Program stars
and observational details are provided in Table \ref{tab:obs}. The
spectra were taken with a resolving power $R=50,000$, except for
HD~196944, for which a $R=90,000$ spectrum was obtained.  Our spectra
cover the wavelength range 3550--5250{\AA}.  Signal-to-Noise (S/N)
ratios per 0.012 {\AA} pixel at 4000 {\AA} are given in the table as
well.  Data reduction was performed in the standard way within the
IRAF\footnote{IRAF is distributed by the National Optical Astronomy
Observatories, which is operated by the Association of Universities
for Research in Astronomy, Inc.  under cooperative agreement with the
National Science Foundation.} environment, following \citet{aoki02b}.

Equivalent widths for isolated lines were measured by fitting Gaussian
profiles to the absorption lines. The line list and equivalent widths
measured are presented in Table~\ref{tab:ew}.


The radial velocity ($V_{\rm r}$) measured for our spectra are also
presented in Table~\ref{tab:obs}. Our separate measurement of the
radial velocity for CS~29526--110 ($V_{\rm r}=201.70\pm
0.26$km~s$^{-1}$ at JD=2452152.7) using WHT/UES shows a variation of
the radial velocity, suggesting binarity for this object. The radial
velocities measured for our Subaru/HDS spectra of CS~22880--027 and
CS~22898--027 agree with the averages of the results by
\citet{preston01} to within 1~km~s$^{-1}$. 
 
\section{Abundance Analysis}\label{ana}

A standard analysis, using the model atmospheres of \citet{kurucz93}
and the equivalent widths measured above, was performed for lines of
\ion{Fe}{1}, \ion{Fe}{2}, and for most of the neutron-capture elements. A
spectrum synthesis technique was applied to the molecular bands of CH,
C$_{2}$, and CN in order to determine the carbon and nitrogen
abundances, which are required for estimates of the effective
temperature.  The line broadening due to macro-turbulence in the stellar
atmospheres, and from the spectrograph itself, was estimated from the clean Fe lines
for each object, and then applied to the spectrum synthesis.


The effective temperatures given in Table \ref{tab:param} were
estimated from broadband colors using previously derived
temperature scales.  For most of our program stars, we applied the
temperature scale based on the de-reddened $(B-V)_0$ color for
carbon-rich, metal-poor stars produced by \citet{aoki02a}. The $B-V$
color of CS~30301--015 is uncertain, hence we assume a larger
uncertainty in effective temperature in the error estimation procedure
described below. The effective temperature of CS~22942--019 was
re-determined after its carbon and nitrogen abundances were estimated,
because abundances for these two elements assumed in the first
determination of the effective temperature were found to be too low,
and the effect of molecular absorption was not included correctly.
$JHK$ photometry data from the interim release of the 2MASS Point
Source Catalog (Skrutskie et al. 1997) are also available for three
stars (CS~29526--110, CS~22898--027 and CS~31062--050).  After correcting
for reddening, the effective temperature scale based on $(V-K)_0$ by
\citet{alonso96} was applied, and yielded consistent results with
those obtained from $(B-V)_0$ colors. The difference between the two
effective temperatures is rather large in CS~29526--110: the effective
temperature estimated from $(V-K)_0$ is higher by 200~K than that from
$(B-V)_0$. We assume a larger uncertainty of $T_{\rm eff}$ for this
object in the error estimates below.

Surface gravities for our program stars were determined from the
ionization balance between \ion{Fe}{1} and \ion{Fe}{2}, the
metallicities were estimated from the abundance analysis for these
lines, and the micro-turbulence was determined from the \ion{Fe}{1}
lines by demanding no dependence of the derived abundance on
equivalent widths. The results are provided in Table \ref{tab:param}.
While the effective temperature of CS~29526--110 is the highest in our
sample, and suggests that this star might be a main-sequence turn-off
star, its gravity ($\log g =3.2$) is not particularly high. This may
be due to non-LTE effects in the analysis of the Fe lines, which were
neglected in the present work.  For the purpose of error estimation,
we therefore assume a larger uncertainty in the derived gravity for
this star.

The carbon abundances were determined using the C$_{2}$ Swan bands for
CS~31062--050, CS~22942--019, and CS~30301--015, stars for which the
CH band at 4323~{\AA} is too strong for abundance analysis. The line
list produced by \citet{aoki97} for C$_{2}$ Swan bands were
applied. For the other five stars, the CH band was used for carbon
abundance determination. The line positions of CH $A-X$ bands were
calculated using the molecular constants derived by
\citet{zachwieja95} and the oscillator strengths of
\citet{brown87}. \citet{aoki02a} showed that the carbon abundance derived
from the C$_{2}$ band is by 0.2~dex higher than that from the CH band for
LP~625--44, indicating that we should include this additional uncertainty of
about 0.2~dex for the carbon abundances. The CN 3883~{\AA} feature was used for
estimates of nitrogen abundances. The line positions of the CN violet system
were calculated using the molecular constants presented by \citet{prasad92} and
the oscillator strengths of \citet{bauschlicher88}. Unfortunately, the nitrogen
abundances of CS~29526--110, CS~22880--074, and CS~30301--015 are unreliable,
because the CN band is either too weak or too strong for abundance
determination.  The results of the carbon and nitrogen abundances are also
given in Table~\ref{tab:param}. The oxygen abundances are assumed to be [O/Fe]
= 0.5 in the analysis, taking account of the well-known overabundances of
oxygen found in metal-poor stars. The derived carbon and nitrogen abundances
are not sensitive to the oxygen abundances assumed in the analysis for warm
($T_{\rm eff}\gtrsim 5000$~K) stars, because the fraction of carbon bound in
the CO molecule is small. The effect of the assumption for the oxygen abundance
most severely appears in CS~30301--015 ($T_{\rm eff}=4750$~K); we tested the
range $0.0<$[O/Fe]$<1.5$ for this star, but found that the effect on abundance
determination for carbon and nitrogen is smaller than 0.1~dex.

The \ion{Pb}{1} 4057~{\AA} line was detected for seven stars in our
sample. Figure~\ref{fig:sp} shows examples of the observed spectra,
and the results of synthetic fits, for HD~196944 and CS~31062--050.
In this analysis we adopted the line data for the hyperfine splitting
and isotope shifts determined by \citet{simons89}, as given in
Table~\ref{tab:pb4057}. We found that the fit between the observed spectra
and the calculated ones is better, when the absolute wavelength of the
Pb lines are slightly shifted. For this reason, we shifted the
wavelength of these lines by 0.005~{\AA}, which is as large as the
uncertainty in the determination of the absolute line position by
\citet{simons89}, in our spectrum synthesis. The CH 4057~{\AA} line
blending with the \ion{Pb}{1} line was calculated for the carbon
abundance which best reproduced other lines of the same CH
band. Therefore, the errors in carbon abundances due to the
uncertainties of oscillator strengths of the band system do not
directly affect the spectrum synthesis for the \ion{Pb}{1} 4057~{\AA}
line.

The effects of hyper-fine splitting and isotope shifts were included,
assuming the solar system isotope ratio for Pb \citep[see ][]{aoki01}.
However, this assumption may not be appropriate for the very metal-poor
stars studied here. For instance, a large production of $^{208}$Pb was
predicted for the {\it s}-process at low metallicity \citep[e.g.,
][]{gallino98}. In order to estimate the uncertainty of the Pb
abundance due to the effect of the isotope ratio assumed, we also
tried the analysis with a single line approximation for the \ion{Pb}{1}
4057~{\AA} line, which simulates the case that $^{208}$Pb is dominant
over the other isotopes. The synthetic spectra calculated with the single
line approximation are shown by dotted lines in
Figure~\ref{fig:sp}. Because of the larger effect of saturation of the
absorption, the derived Pb abundance from the single line
approximation is larger than that from the analysis including isotope
shifts with the assumption of the solar system isotope ratio. The
difference of the resulting abundance is about 0.2~dex for
CS~31062--050, in which the \ion{Pb}{1} 4057~{\AA} line is strongest
among our objects. The effect of the isotope ratios assumed in the
analysis for other objects is about 0.1~dex or smaller. We include
these uncertainties in the error estimates below.



We found that the abundances of the lighter neutron-capture elements
(e.g., Ba), as well as the stellar parameters, of CS~31062--050 are
quite similar to those of the well-studied object LP~625--44.  However,
the Pb absorption line in CS~31062--050 is significantly stronger than
that of LP~625--44 \citep[see Figure 6 of ][]{aoki02b}. This clearly
indicates a large dispersion in the ratio of the abundances between
the lighter {\it s}-process elements and Pb.

The \ion{Pb}{1} 3683~{\AA} line was also detected in five of the above seven
stars. Though the quality of the spectra at 3683~{\AA} is worse than that at
4057~{\AA}, we found that the abundances derived from the 3683~{\AA} line agree
well with those obtained from the 4057~{\AA} line within the errors. We note
that the abundances based on the 4057~{\AA} line were adopted as the final
result given in Table~\ref{tab:res}; those based on the  3683~{\AA} line 
were only used for confirmation. An upper limit on the Pb abundance was
obtained for CS~22942--019, in which the above Pb lines were not detected.

The abundances of most of the other neutron-capture elements were determined by
standard analysis based on the measured equivalent widths, while the abundances
of Ba and Eu were determined from spectrum synthesis. The effects of hyper-fine
splitting and isotope shifts were included in the analysis of Ba, La, Nd, and
Eu lines using the line lists of \citet{mcwilliam98} and \citet{aoki01}.

In order to estimate the random error in the analysis, we first
calculated the dispersion of the abundances for individual \ion{Fe}{1}
lines in each star.  In addition, we assumed the random error in {\it
gf}-values to be 0.1~dex, and added it in quadrature to the above
dispersion. The random error was evaluated by dividing this estimate by
$n^{1/2}$ ($n$ is the number of lines used in the analysis) for each
neutron-capture element. For Pb abundances, we also included the
uncertainties due to the isotope ratio assumed in the spectrum
synthesis (0.2~dex for CS~31062--050, and 0.1~dex for other stars). The
errors in the abundance determination from the uncertainties of the
atmospheric parameters were evaluated for $\sigma (T_{\rm eff})=100$K,
$\sigma (\log g)=0.3$, and $\sigma (v)=0.5~$km s$^{-1}$ for HD~196944 and
CS~22898--027. These results were then applied to four of the remaining
stars.  We assumed $\sigma (T_{\rm eff})=200$~K for CS~29526--110 and
CS~30301--015, and $\sigma (\log g)=0.5$ for CS~29526--110, taking the
uncertainties noted above into consideration. We finally derived the
total uncertainty by adding in quadrature the individual errors, and
list them in Table \ref{tab:res}.

We also tried to estimate the effective temperatures for seven stars
in our sample based on the Balmer line strengths measured from the
medium-resolution spectra obtained during the HK survey follow-up. These
effective temperatures are systematically lower, by about 300~K, than
those derived above from the broadband colors. The effects of the changes of
effective temperature by $-300$K on the resulting [Fe/H] and [Pb/Ba], which
will be discussed below, are about $-0.20$ and $+0.25$~dex, respectively.
Though these effects are important in the determination of the abundances, and
will be investigated more thoroughly in our future work, the systematic shift
of the abundance ratios does not change the following discussion on the
dispersion in the values of [Pb/Ba] found in our sample.

The abundances of HD~196944 derived in this work generally agree with those by
\citet{zacs98} for iron and carbon, and \citet{vaneck01} for other elements,
including Pb, within 0.2~dex, except for La (our result is 0.38~dex higher
than that of \citet{vaneck01}). The agreement between the abundances of
HD~196944 determined by \citet{vaneck01} and those obtained in the present
analysis (following the previous work for LP~625--44) indicates that the
significant difference in the abundance ratios of Pb, as compared to the
lighter {\it s}-process elements, in the objects analyzed by \citet{vaneck01}
and those by \citet{aoki01} is not due to differences in the respective
analysis procedures.

\section{Discussion and Concluding Remarks}\label{sec:disc}

The resulting abundances for two of the stars in our sample are shown
in Figure~\ref{fig:abund}, along with the scaled abundance patterns of
solar system material, the main {\it s}-process component, and the
(inferred) {\it r}-process pattern \citep{arlandini99}. These
comparisons are useful to distinguish the origins of the
neutron-capture elements detected in our objects.  The abundance
patterns of elements with $56\leq Z\leq 63$ in our stars agree with
the {\it s}-process pattern much better than with the {\it r}-process
pattern. This result implies that, as expected, the neutron-capture
elements in our stars principally originate in the {\it s}-process.
The values of [Ba/Eu] for four of the stars in our sample
(CS~29526--110, CS~22898--027, CS~31062--012, and CS~31062--050) are
significantly lower ([Ba/Eu]=0.36$\sim$0.47) than seen in the main
{\it s}-process component \citep[{[Ba/Eu]=+1.15};][]{arlandini99}.  We
suggest, however, that the derived abundance ratios of [Ba/Eu] have
been produced by an {\it s}-process which produces {\it different} abundance
ratios from that of the main {\it s}-process component. Indeed,
\citet{goriely00} predicted lower values of [Ba/Eu] ($\sim 0.4$) for
yields of metal-deficient AGB stars, though the [Ba/Eu] value of our
objects is not correlated with metallicity. To explain these low
values of [Ba/Eu] by the mixture of the abundance ratios of the {\it
r}- and {\it s}-process components in the solar system, we must assume
that about 80-90\% of Eu nuclei were produced by the {\it r}-process.
If this were true, it follows that these four stars show very large
excesses of their {\it r}-process elements ([Eu/Fe]=1.5$\sim$1.8),
similar to CS~31082--001, an extreme {\it r}-process-element-enhanced
star (Cayrel et al. 2001; Hill et al. 2002).  However, recent
high-resolution studies of metal-poor giants indicate that stars with
such large excesses of {\it r}-process elements are quite rare.  An
ongoing survey by Christlieb et al. (2002, in preparation) confirms
the original suggestion by Beers (private communication) that roughly
3\% of giants with [Fe/H] $< -2.5$ exhibit [{\it r}/Fe]$ \ge
+1.0$. Hence, it would be difficult to appeal to this explanation to
account for the excesses of Eu in the four stars in our sample. We
treat our stars as having the abundance patterns produced by the {\it
s}-process for neutron-capture elements in this paper, because the
elements discussed below (Ba and Pb) should principally originate from the
{\it s}-process, even though {\it r}-process contamination may have
contributed to a portion of the observed Eu excesses.

Figure~\ref{fig:bapb} shows the ratio [Pb/Ba], representing the ratio
of abundances between elements located at the second and third {\it
s}-process peaks, as a function of metallicity ([Fe/H]).  There exists
a large scatter in [Pb/Ba] for these stars, larger than can be
accounted for by the errors in the abundance analysis.  It is not
apparent that our stars can be readily classified into separate groups
on the basis of the present data. There may be evidence for an
decreasing trend of [Pb/Ba] with decreasing [Fe/H] (the correlation
estimated from objects in Figure 3, excluding CS~22942--019, is
represented as [Pb/Ba]$=2.62+0.88$[Fe/H] with the standard error of
the slope of 0.47, i.e., the possible slope is significant at the 1.9
sigma level). Additional data will presumably help to clarify these
questions. Additional studies of {\it s}-process nucleosynthesis at low
metallicity should be carried out to explore the possible reasons
for this large dispersion, and the
possible dependence on metallicity, of the Pb/Ba ratios.

Comparison of these results with the predictions of AGB
nucleosynthesis models of \citet{gallino98} and \citet{busso99}
provides a constraint on their $^{13}$C pocket models\footnote{
Uncertainties in the amount of $^{13}$C in the pocket reflects the
unknown amount of mixing of protons from the H-rich envelope down into
the C-rich zone.  Recent model calculations by Cristallo et al. (2001;
see also Goriely \& Mowlavi 2000) treat the proton mixing as an input
parameter, rather than the amount of $^{13}$C in the ``pocket''.  The
efficiency of the $^{13}$C as a neutron source for the {\it s}-process is also
affected by the profile of neutron poisons like $^{14}$N.
}. Figure 1 of \citet{ryan01} shows the metallicity
dependence of [Pb/Fe] and [Ba/Fe] predicted by the model for
1.5~$M_{\odot}$ AGB stars, compared with the abundance ratios observed
in LP~625--44.  The value of [Pb/Ba] is sensitive to the adopted
$^{13}$C profile, represented by the normalization factor of the
$^{13}$C pocket in the standard model of \citet{busso99}.  The [Pb/Ba]
values found in our stars with $-2.7 \leq$[Fe/H]$\leq -1.9$ distribute
from $-0.3$ to +1.2 (Figure~\ref{fig:bapb}). This range of the
abundance ratios is not explained by the standard model, which
predicts [Pb/Ba]$\sim +2$, but can be explained by models where the  
amount of $^{13}$C produced as the neutron source for the $s$-process
is lowered by factors of 6-24 relative to the standard model.

We would also like to mention the model of the {\it s}-process in AGB
stars recently proposed by \citet{iwamoto02}. Their model for AGB
stars with [Fe/H] = $-2.7$ and $M=2M_{\odot}$ shows a possible
production of $^{13}$C, a candidate for the neutron source of the {\it
s}-process, as the result of proton mixing into the hot He shell at
the second thermal pulse \citep[see also ][]{fujimoto00}. Since the
value of neutron exposure predicted by their model is rather high
($\tau \sim 1~$mb$^{-1}$), one may expect high Pb abundances relative
to those of lighter neutron-capture elements. However, the small
number of the exposures (only one in their model) is preferable to account for
the low Pb abundances that we find in some stars in our sample
\citep{aoki01}. This mechanism for producing the required free
neutrons is expected to occur only in very metal-poor ([Fe/H]$\lesssim
-2.5$) AGB stars, and may explain the abundances of some objects with
lower metallicity and high [Pb/Ba] values.  The comparison with this
model is, however, still quite speculative, and further theoretical
study for the possibility of flash-driven proton mixing and its
contribution to the {\it s}-process is required.

Finally, we point out that clear temporal variations of radial
velocity, which directly implies the binarity of the object, have thus
far been detected for only three of the carbon-enhanced, metal-poor stars we
have been studying in recent years:  CS~22942--019 \citep{preston01},
CS~29526--110 (section \ref{sec:obs}), and LP~625--44 \citep{aoki00}.
\citet{preston01} reported that {\it none} of the three carbon-rich,
metal-poor subgiants in their sample, including the stars CS~22898--027
and CS~22880--074 studied here, exhibited radial velocity variations
exceeding 0.5~kms$^{-1}$ over an 8 year period.  Furthermore, no
radial velocity variation has been found for the subgiant (or dwarf)
LP~706--7 \citep{norris97}. These results suggest that the enhancement
of neutron-capture elements in these subgiants may not be explained as
a result of the transfer of material rich in {\it s}-process elements
across a binary system including an AGB star. For comparisons with the
predictions of AGB models, confirmation of binarity based on the
radial velocity monitoring is strongly desired.  It is interesting
to note the large dispersion
of [Pb/Ba] values of the three objects (CS~22942--019,
CS~29526--110, and LP~625--44) for which the variation of radial
velocity have been observed.  This suggests that the
Pb/Ba ratios produced by {\it s}-process nucleosynthesis in
metal-deficient AGB stars show a large scatter, although the sample is
still too small to be confident in this regard. 

The present work clearly shows a variety in the abundance ratios of
neutron-capture elements for carbon-rich, metal-poor stars, and underscores the
importance of further studies for individual objects, including long-term
radial velocity monitoring, for development of a better understanding of heavy
element production by the {\it s}-process operating at low metallicity.

\acknowledgments 

The authors are pleased to acknowledge valuable discussions with Drs.
R. Gallino and M. Busso on the variation of {\it s}-process
nucleosynthesis in their Population II AGB models. We would also like
to thank the anonymous referee for useful comments for improving this
paper.  T.C.B. acknowledges partial support for this work from grant
AST00-98549, awarded by the U.S. National Science Foundation.


\begin{deluxetable}{p{23mm}cccp{5mm}p{10mm}p{4mm}cl}
\tablewidth{0pt}
\tablecaption{PROGRAM STARS AND OBSERVATIONS \label{tab:obs}}
\startdata
\tableline
\tableline
Star & $V$ & $B-V$ & $E(B-V)$ & $GP$ & Exp.$^{\rm a}$ & S/N & $V_{\rm r}$ (kms$^{-1}$) & Obs. Date (JD)\\ 
\tableline
CS22942--019 & 12.71 & 0.86 & 0.02 & 7.60 &  60 (2) &  61 & $-$232.88$\pm$0.40 & 20 Aug., 2000 (2451778) \\
CS29526--110 & 13.37 & 0.35 & 0.00 & 1.94 &  90 (3) &  62 & +208.35$\pm$0.76 & 28 Jan., 2001 (2451938) \\
CS22898--027 & 12.76 & 0.50 & 0.03 & 4.71 &  60 (3) &  49 &  $-$47.90$\pm$0.47 & 22 July, 2001 (2452113) \\
HD~196944    &  8.41 & 0.61 & 0.00 & -    &  20 (1) & 150 & $-$174.76$\pm$0.36 & 25 July, 2001 (2452116) \\
CS31062--012 & 12.10 & 0.46 & 0.00 & 3.67 &  25 (1) &  57 &  +80.76$\pm$0.37 & 25 July, 2001 (2452117) \\
CS31062--050 & 13.04 & 0.64 & 0.02 & 6.41 &  50 (2) &  47 &   +7.75$\pm$0.67 & 25 July, 2001 (2452117) \\
CS30301--015 &     - & 0.99:& -    & 7.88 &  50 (2) &  53 &  +86.46$\pm$0.43 & 25 July, 2001 (2452116) \\
CS22880--074 & 13.27 & 0.57 & 0.04 & 5.71 &  60 (2) &  42 &  +59.57$\pm$0.48 & 25 July, 2001 (2452116) \\
\tableline
\enddata
~ \\

$^{\rm a}$ Exposure time (minute) and number of exposures.

\end{deluxetable}

\clearpage

\begin{deluxetable}{ccccccccccc}
\tablewidth{0pt}
\tablecaption{EQUIVALENT WIDTHS \label{tab:ew}}
\startdata
\tableline
\tableline
 Wavelength & LEP  & $\log gf$ & 29526 & 22898 & 31062 & 22880 & 31062 & HD~196944 & 22942 & 30301 \\
   ({\AA})  & (eV) &           & --110 & --027 & --012 & --074 & --050 &           & --019 & --015 \\ 
\tableline
     Sr II&&&&&&&&&& \\ 
   4077.71&    0.00&    0.15&   102.3&   115.4&    76.6&   127.6&   146.9&   179.7& 488:$^{\dag}$ & 176:$^{\dag}$  \\
   4161.80&    2.94&   $-$0.50&    ... &    ... &    ... &    ... &    ... &    16.1&    ... &    ...  \\
   4215.52&    0.00&   $-$0.17&    90.3&    97.8&    56.2&   104.6&   127.2&   184.7& 364:$^{\dag}$ &    ...  \\
      Y II&&&&&&&&&& \\ 
   3600.73&    0.18&    0.28&    ... &    48.6&    ... &    ... &    ... &    95.4$^{\ast}$&    ... &    ...  \\
   3611.04&    0.13&    0.01&    ... &    ... &    ... &    34.2&    ... &    86.0$^{\ast}$&    ... &    ...  \\
   3628.70&    0.13&   $-$0.71$^{\rm a}$&    ... &    ... &    ... &    ... &    ... &    58.7$^{\ast}$&    ... &    ...  \\
   3788.69&    0.10&   $-$0.06&    ... &    ... &    ... &    ... &    ... &    96.8&    ... &    ...  \\
   3950.35&    0.10&   $-$0.49&    ... &    25.0&     7.6&    30.2&    ... &    78.7&   101.0&    ...  \\
   4682.32&    0.41&   $-$1.51$^{\rm a}$&    ... &    ... &    ... &    ... &    ... &    22.8&    ... &    ...  \\
   4883.68&    1.08&    0.07&    ... &    ... &    ... &    ... &    ... &    56.2&   103.1&    61.2 \\
   5087.42&    1.08&   $-$0.16&    ... &    ... &    ... &    ... &    ... &    42.7&    ... &    55.2 \\
   5200.41&    0.99&   $-$0.57&    ... &    ... &    ... &    ... &    ... &    36.6&    ... &    33.0 \\
   5205.72&    1.03&   $-$0.34&    ... &    ... &    ... &    ... &    ... &    35.7&    ... &    41.8 \\
      Zr II&&&&&&&&&& \\ 
   3573.08&    0.32&   $-$1.04&    ... &    ... &    ... &    ... &    ... &    33.5&    ... &    ...  \\
   3630.02&    0.35&   $-$1.11&    ... &    ... &    ... &    ... &    ... &    25.1&    ... &    ...  \\
   3998.97&    0.56&   $-$0.67&    ... &    ... &    ... &    ... &    ... &    51.2&    75.4&    ...  \\
   4029.68&    0.71&   $-$0.76&    ... &     8.0&    ... &    ... &    23.6&    41.9&    ... &    ...  \\
   4050.33&    0.71&   $-$1.00&    ... &    ... &    ... &    ... &    ... &    ... &    59.2&    ...  \\
   4149.20&    0.80&   $-$0.03&    28.7&    ... &    ... &    ... &    ... &    ... &    ... &    ...  \\
   4208.98&    0.71&   $-$0.46&    ... &    16.9&    ... &    ... &    31.9&    47.0&    ... &    ...  \\
   4258.05&    0.56&   $-$1.13&    ... &    ... &    ... &    ... &    ... &    29.9&    ... &    ...  \\
   4443.00&    1.49&   $-$0.33&    ... &    ... &    ... &    ... &    ... &    ... &    56.4&    ...  \\
      Ba II&&&&&&&&&& \\ 
   3891.78&    2.51&    0.28$^{\rm b}$&    ... &    49.5&    ... &    ... &    ... &    34.0& 50.7$^{\dag}$ &    ...  \\
   4130.65&    2.72&    0.56$^{\rm c}$&    50.2&    64.3&    23.1& 31.0$^{\dag}$& 85.2&    49.1& 69.4$^{\dag}$ &    81.7 \\
   4166.00&    2.72& $-$0.41     &12.7$^{\dag}$ &    15.1&    ... &    ... &    27.7&    11.5& 22.1$^{\dag} $ &    ...  \\
      La II&&&&&&&&&& \\ 
   3790.83&    0.13&    0.03&    ... &    ... &    ... &    ... &    70.7&    ... &    ... &    ...  \\
   3949.10&    0.40&    0.49&    69.0&   107.9&    34.4&    50.4&129.1$^{\ast}$&    95.1&    ... &    ...  \\
   3988.52&    0.40&    0.21&    44.3&    ... &    ... &    33.3&120.9$^{\ast}$&    67.9&    ... &    70.3 \\
   3995.75&    0.17&   $-$0.06&    34.3&    59.5&    18.6&    ... & 89.4$^{\ast}$&    54.2&    60.4&    73.1 \\
   4086.72&    0.00&   $-$0.07&    ... &    54.3&    ... &    33.0&    73.7&    ... &    61.1&    ...  \\
   4123.23&    0.32&    0.13&    40.3&    ... &    19.1&    ... &    74.2&    50.9&    ... &    ...  \\
   4322.51&    0.17&   $-$0.93&    ... &    20.7&    ... &    ... &    51.7&    15.8&    ... &    ...  \\
   4333.74&    0.17&   $-$0.06&    54.1&    ... &    ... &    ... &    ... &    ... &    ... &    ...  \\
   4429.90&    0.23&   $-$0.35&    36.5&    ... &    ... &    ... &    ... &    47.6&    ... &    ...  \\
   4662.51&    0.00&   $-$1.24&    ... &    14.4&    ... &    ... &    ... &    ... &    ... &    32.7 \\
      Ce II&&&&&&&&&& \\ 
   3853.16& 0.00&   $-$0.36$^{\rm d}$&    ... &    21.5&    ... &    ... &    ... &    ... &    ... &    ...  \\
   3909.31&    0.45&   $-$0.52&    ... &    ... &    ... &    ... &    19.3&     9.0&    ... &    ...  \\
   3940.34& 0.32&   $-$0.20$^{\rm d}$&    ... &    16.6&    ... &    ... &    25.8&    24.7&    ... &    30.1 \\
   3942.75& 0.86&    0.41$^{\rm d}$&    ... &    24.8&    ... &    ... &    ... &    23.9&    ... &    33.7 \\
   3960.91&    0.32&   $-$0.20&    ... &    ... &    ... &    ... &    ... &    15.4&    ... &    ...  \\
   3980.88&    0.71&    0.03&    ... &    ... &    ... &    ... &    ... &    12.0&    ... &    ...  \\
   3984.68&    0.96&    0.41&    ... &    ... &    ... &    ... &    ... &    15.2&    ... &    ...  \\
   3992.39& 0.45&   $-$0.13$^{\rm d}$&    ... &    10.4&    ... &    ... &    ... &    ... &    ... &    ...  \\
   3999.24& 0.29&    0.32$^{\rm d}$&    ... &    ... &    8.5 &    17.5&    45.6&    32.0&    39.8&    53.0 \\
   4014.90& 0.53&    0.05$^{\rm d}$&    ... &    12.3&    ... &    ... &    34.8&    ... &    ... &    ...  \\
   4046.34& 0.55&   $-$0.11$^{\rm d}$&    ... &     8.9&    ... &    ... &    ... &    15.8&    ... &    ...  \\
   4053.51& 0.00&   $-$0.60$^{\rm d}$&    ... &    ... &    ... &    ... &    26.1&    17.7&    23.8&    ...  \\
   4068.84& 0.70&   $-$0.13$^{\rm d}$&    ... &    ... &    ... &    ... &    ... &    16.6&    ... &    ...  \\
   4072.92&    0.33&   $-$0.71&    ... &    11.0&    ... &    ... &    ... &    14.7&    ... &    ...  \\
   4073.48& 0.48&    0.28$^{\rm d}$&    ... &    27.2&    ... &    19.9&    35.6&    31.7&    ... &    ...  \\
   4118.14& 0.70&    0.10$^{\rm d}$&    ... &    ... &    ... &    ... &    37.9&    23.9&    ... &    ...  \\
   4120.83& 0.32&   $-$0.51$^{\rm d}$&    ... &    ... &    ... &    ... &    ... &    19.4&    ... &    ...  \\
   4124.79& 0.68&   $-$0.10$^{\rm d}$&    ... &    ... &    ... &    ... &    30.7&    ... &    ... &    ...  \\
   4127.37& 0.68&    0.19$^{\rm d}$&    ... &    ... &    ... &    ... &    37.1&    ... &    ... &    ...  \\
   4137.65& 0.52&    0.32$^{\rm d}$&    ... &    34.8&    11.3&    ... &    49.9&    39.2&    ... &    59.3 \\
   4145.00& 0.70&    0.03$^{\rm d}$&    ... &    ... &    ... &    ... &    ... &    14.6&    20.5&    ...  \\
   4165.61& 0.91&    0.51$^{\rm d}$&    ... &    22.8&    ... &    ... &    38.4&    28.8&    ... &    ...  \\
   4185.33&    0.42&   $-$0.56&    ... &    ... &    ... &    ... &    ... &    10.6&    ... &    ...  \\
   4193.87& 0.55&   $-$0.40$^{\rm d}$&    ... &    ... &    ... &    ... &    18.5&    14.0&    ... &    ...  \\
   4222.60& 0.12&   $-$0.21$^{\rm d}$&    ... &    28.9&     8.8&    ... &    44.6&    33.5&    ... &    58.1 \\
   4349.79& 0.53&   $-$0.19$^{\rm d}$&    ... &    ... &    ... &    ... &    25.6&    ... &    ... &    ...  \\
   4418.78& 0.86&    0.26$^{\rm d}$&    12.7&    23.9&    ... &    ... &    ... &    30.4&    ... &    ...  \\
   4449.34& 0.61&   $-$0.11$^{\rm d}$&    ... &    21.6&    ... &    ... &    ... &    25.2&    ... &    ...  \\
   4460.21& 0.48&    0.25$^{\rm d}$&    ... &    35.1&    ... &    ... &    ... &    40.0&    ... &    77.8 \\
   4471.24& 0.70&    0.23$^{\rm d}$&    ... &    ... &    10.6&    ... &    44.9&    33.4&    ... &    ...  \\
   4483.90&    0.86&    0.09&    ... &    ... &    ... &    ... &    ... &    ... &    47.1&    ...  \\
   4486.91& 0.29&   $-$0.39$^{\rm d}$&    ... &    18.6&    ... &    ... &    39.7&    31.7&    35.5&    46.1 \\
   4551.30&    0.74&   $-$0.49&    ... &    ... &    ... &    ... &    19.2&    ... &    ... &    24.5 \\
   4560.28& 0.91&    0.08$^{\rm d}$&    ... &    17.4&    ... &    ... &    37.3&    19.9&    ... &    53.6 \\
   4560.96&    0.68&   $-$0.47&    ... &    ... &    ... &    ... &    22.7&     9.2&    ... &    23.7 \\
   4562.36& 0.48&    0.16$^{\rm d}$&    ... &    ... &     9.9&    18.8&    57.5&    36.0&    49.0&    70.5 \\
   4593.93& 0.70&   $-$0.03$^{\rm d}$&    ... &    19.0&    11.9&    ... &    49.6&    19.9&    ... &    55.8 \\
   4628.16& 0.52&    0.09$^{\rm d}$&    ... &    18.6&    10.2&    ... &    48.5&    29.2&    ... &    57.4 \\
   5187.45& 1.21&    0.16$^{\rm d}$&    ... &    ... &    ... &    ... &    ... &    ... &    ... &    14.2 \\
     Nd II&&&&&&&&&& \\ 
   3780.40&    0.47&   $-$0.27&    ... &    ... &    ... &    ... &    ... &    25.2&    ... &    ...  \\
   3784.25&    0.38&    0.23&    ... &    34.1&     7.0&    ... &    ... &    30.7&    28.8&    ...  \\
   3911.17&    0.47&    0.20&    ... &    ... &    ... &    ... &    52.2&    27.6&    ... &    ...  \\
   3927.10&    0.18&   $-$0.52&    ... &    ... &    ... &    ... &    33.0&    ... &    18.2&    ...  \\
   3941.51&    0.15&    0.06&    ... &    ... &    14.8&    ... &    ... &    36.5&    ... &    ...  \\
   3979.49&    0.20&   $-$0.11&    ... &    ... &    ... &    ... &    ... &    29.5&    30.1&    ...  \\
   3991.74&    0.00&   $-$0.40&    ... &    28.9&    ... &    ... &    54.8&    33.0&    50.8&    ...  \\
   4012.70&    0.00&   $-$0.64&    ... &    ... &    ... &    ... &    36.3&    18.0&    20.8&    ...  \\
   4020.87&    0.32&   $-$0.18&    ... &    ... &    ... &    ... &    47.1&    29.1&    47.1&    ...  \\
   4021.34&    0.32&    0.24&    ... &    ... &     7.3&    ... &    43.2&    21.3&    ... &    ...  \\
   4023.00&    0.20&   $-$0.19&    ... &    23.4&    ... &    ... &    50.8&    18.5&    ... &    ...  \\
   4038.12&    0.18&   $-$0.78&    ... &    ... &    ... &    ... &    44.4&    ... &    ... &    ...  \\
   4051.15&    0.38&   $-$0.21&    ... &    19.9&    ... &    ... &    34.5&    ... &    ... &    ...  \\
   4059.96&    0.20&   $-$0.33&    ... &    ... &    ... &    ... &    ... &    ... &    ... &    17.5 \\
   4061.09&    0.32&   $-$0.34&    ... &    ... &    ... &    ... &    ... &    48.0&    ... &    ...  \\
   4069.28&    0.06&   $-$0.33&    ... &    22.7&    ... &    ... &    39.1&    21.1&    16.0&    43.3 \\
   4075.12&    0.20&   $-$0.40$^{\rm e}$&    11.6&    35.6$^{\ast}$&    22.4$^{\ast}$&    ... &    ... &    ... &    ... &    ...  \\
   4116.77&    0.06&   $-$1.65&    ... &    ... &    ... &    ... &    24.7&    ... &    ... &    ...  \\
   4133.36&    0.32&   $-$0.34&    ... &    ... &    ... &    ... &    ... &    17.8&    ... &    ...  \\
   4156.08&    0.18&    0.10$^{\rm f}$&    28.4&    43.2$^{\ast}$&    ... &    ... &    ... &    ... &    ... &    ...  \\
   4160.57&    0.56&   $-$0.45&    ... &    ... &    ... &    ... &    25.6&    ... &    ... &    ...  \\
   4177.32&    0.06&   $-$0.04$^{\rm e}$&    ... &    38.7$^{\ast}$&    ... &    22.5&    61.7$^{\ast}$&    ... &    ... &    ...  \\
   4358.17&    0.32&   $-$0.21&    ... &    30.9&    ... &    ... &    ... &    ... &    ... &    ...  \\
   4446.39&    0.20&   $-$0.63&    ... &    19.0&    ... &    ... &    44.9&    29.7&    ... &    37.8 \\
   4462.99&    0.56&   $-$0.07&    ... &    35.4&    10.6&    ... &    56.9&    28.5&    ... &    ...  \\
   4516.36&    0.32&   $-$0.75&    ... &    ... &    ... &    ... &    26.1&     8.4&    ... &    ...  \\
     Sm II&&&&&&&&&& \\ 
   3979.20& 0.54&   $-$0.19$^{\rm g}$&    ... &    ... &    ... &    ... &    ... &     7.1&    19.6&    13.8 \\
   4064.58& 0.33&   $-$0.27$^{\rm g}$&    ... &    ... &    ... &    ... &    19.5&    ... &    ... &    ...  \\
   4244.70& 0.28&   $-$0.73$^{\rm h}$&    ... &    ... &    ... &    ... &    14.4&    ... &    ... &    ...  \\
   4318.94&    0.28&   $-$0.27&    ... &    11.7&    ... &    ... &    27.6&    15.1&    ... &    ...  \\
   4424.34& 0.49&    0.07$^{\rm g}$&    ... &    ... &    ... &    ... &    ... &    11.9&    ... &    28.0 \\
   4434.32&    0.38&   $-$0.26&    ... &    16.0&    ... &    ... &    33.8&    14.7&    ... &    ...  \\
   4467.34& 0.66&    0.12$^{\rm i}$&    ... &    13.1&    ... &    ... &    30.4&    12.5&    ... &    27.2 \\
   4499.48& 0.25&   $-$1.00$^{\rm h}$&    ... &    ... &    ... &    ... &    23.6&    ... &    ... &    ...  \\
   4537.95&    0.49&   $-$0.23&    ... &    ... &    ... &    ... &    30.5&    ... &    ... &    ...  \\
   4642.24& 0.38&   $-$0.52$^{\rm h}$&    ... &    ... &    ... &    ... &    33.1&    ... &    ... &    ...  \\
     Eu II&&&&&&&&&& \\ 
   3819.67& 0.00& 0.49$^{\rm j}$& 55.6$^{\dag}$ & 83.5$^{\dag}$ & 19.1$^{\dag}$ &    ... & 132.4$^{\dag}$ & 43.4$^{\dag}$ & 60.7$^{\dag}$ &    ...  \\
   3907.10      & 0.21& 0.20&    ...           &    ...       &    ...       &    ... & 87.4$^{\dag}$ & 15.4$^{\dag}$ &    ... &    ...  \\
   4129.70& 0.00& 0.20$^{\rm j}$& 44.6$^{\dag}$ & 55.7$^{\dag}$ & 12.8$^{\dag}$ &    ... & 116.0$^{\dag}$ & 22.4$^{\dag}$ & 40.5$^{\dag}$ & 47.1$^{\dag}$  \\
   4205.05& 0.00& 0.12$^{\rm j}$& 33.7$^{\dag}$ & 48.8$^{\dag}$ & 8.7$^{\dag}$  & 9.8$^{\dag}$ &    ... & 21.1$^{\dag}$ & 35.0$^{\dag}$ &    ...  \\
     Dy II&&&&&&&&&& \\ 
   3944.68&    0.00&    0.08&    ... &    28.3&    ... &    11.9&    ... &    32.7&    34.9&    48.1 \\
   4077.96&    0.10&   $-$0.03&    ... &    31.0&    ... &    ... &    53.4&    ... &    ... &    ...  \\
     Er II&&&&&&&&&& \\ 
   3616.56&    0.00&   $-$0.33$^{\rm k}$&    ... &    27.9&    ... &    19.0&    ... &    23.4&    ... &    ...  \\
   3786.84&    0.00&   $-$0.64&    ... &    27.4&    ... &    ... &    ... &    ... &    ... &    ...  \\
   3830.48&    0.00&   $-$0.37$^{\rm k}$&    ... &    43.7&    ... &    ... &    ... &    ... &    ... &    ...  \\
   3896.23&    0.05&   $-$0.24$^{\rm k}$&    ... &    36.5&    ... &    ... &    ... &    29.5&    ... &    ...  \\
     Pb I&&&&&&&&&& \\ 
   3683.46&    0.97&   $-$0.52&    30.1$^{\dag}$&    24.2$^{\dag}$&    ... &    ... &    45.3$^{\dag}$&    24.1$^{\dag}$&    ... &    26.2$^{\dag}$ \\
   4057.82&    1.32&   $-$0.20&    34.9$^{\dag}$&    27.4$^{\dag}$&     6.4$^{\dag}$&    15.4$^{\dag}$&    53.3$^{\dag}$&    24.5$^{\dag}$&    ... &    20.7$^{\dag}$ \\
\tableline
\enddata

$\ast$ Lines which were not used in the analysis.

$\dag$ Synthesized value calculated for the abundance derived by spectrum synthesis.

References for $gf$--values.--- $^{\rm a}$\citet{HLGBW82}, $^{\rm b}$\citet{sneden96},
$^{\rm c}$\citet{mcwilliam98}, $^{\rm d}$ \citet{CB62} +0.23~dex, $^{\rm e}$
\citet{CC83}, $^{\rm f}$ \citet{WM80}, $^{\rm g}$ \citet{CB62}+0.49~dex,
$^{\rm h}$ \citet{BGHL89}, $^{\rm i}$ \citet{vogel88}, $^{\rm j}$
\citet{biemont82}, $^{\rm k}$ \citet{ML83}, others: \citet{aoki02b} or \citet{aoki01}

\end{deluxetable}

\clearpage
\begin{deluxetable}{lcccccc}
\tablewidth{0pt}
\tablecaption{STELLAR PARAMETERS \label{tab:param}}
\startdata
\tableline
\tableline
Star & $T_{\rm eff}$(K)$^{\ast}$ & $\log g$ & $v_{\rm tur}$(kms$^{\small -1}$) & [Fe/H] & [C/Fe] & [N/Fe] \\ 
\tableline
CS~29526--110 & 6500$^{\rm a ~ ~ }$ & 3.2 & 1.6 & $-$2.38 & 2.2 & 1.4: \\
CS~22898--027 & 6250$^{\rm a,b}$ & 3.7 & 1.5 & $-$2.25 & 2.2 & 0.9 \\
CS~31062--012 & 6250$^{\rm a ~ ~ }$ & 4.5 & 1.5 & $-$2.55 & 2.1 & 1.2 \\
CS~22880--074 & 5850$^{\rm a ~ ~ }$ & 3.8 & 1.4 & $-$1.93 & 1.3 & $-$0.1: \\
CS~31062--050 & 5600$^{\rm a,b}$ & 3.0 & 1.3 & $-$2.32 & 2.0 & 1.2 \\
HD~196944    & 5250$^{\rm a ~ ~ }$ & 1.8 & 1.7 & $-$2.25 & 1.2 & 1.3 \\
CS~22942--019 & 5000$^{\rm a ~ ~ }$ & 2.4 & 2.1 & $-$2.64 & 2.0 & 0.3 \\
CS~30301--015 & 4750$^{\rm a ~ ~ }$ & 0.8 & 2.2 & $-$2.64 & 1.6 & 0.6: \\
\tableline
\enddata
~ \\

$\ast$ The effective temperature is determined based on the 
following color indices: a:$(B-V)_{0}$ and b:$(V-K)_{0}$. 
\end{deluxetable}

\begin{deluxetable}{ccc}
\tablewidth{0pt}
\tablecaption{Hyperfine and isotope splitting of Pb~4057.8~{\AA} line \label{tab:pb4057}}
\startdata
\tableline
\tableline
Wavelength ({\AA}) & fraction & isotope \\ 
\tableline
4057.7524 & 0.015  & 207b \\
4057.7895 & 0.135  & 207a \\ 
4057.7895 & 0.524  & 208 \\
4057.8030 & 0.236  & 206  \\
4057.8150 & 0.015  & 204  \\
4057.8251 & 0.075  & 207c \\ 
\tableline
\enddata
\end{deluxetable}

\begin{deluxetable}{@{}l@{\extracolsep{\fill}}c@{\extracolsep{\fill}}c@{\extracolsep{\fill}}c@{\extracolsep{\fill}}c@{\extracolsep{\fill}}c@{\extracolsep{\fill}}c@{\extracolsep{\fill}}c@{\extracolsep{\fill}}c@{\extracolsep{\fill}}c@{\extracolsep{\fill}}c@{\extracolsep{\fill}}c@{\extracolsep{\fill}}c@{\extracolsep{\fill}}c@{\extracolsep{\fill}}c@{\extracolsep{\fill}}c@{\extracolsep{\fill}}c@{\extracolsep{\fill}}c@{\extracolsep{\fill}}c@{\extracolsep{\fill}}c}
\tablecaption{[FE/H] AND RELATIVE ABUNDANCE, [X/FE] \label{tab:res}}
\startdata
\tableline
\tableline
 &\multicolumn{4}{c}{CS~29526--110} &&\multicolumn{4}{c}{CS~22898--027} && \multicolumn{4}{c}{CS~31062--012} && \multicolumn{4}{c}{CS~22880--074}  \\ 
  \cline{2-5}  \cline{7-10}    \cline{12-15}   \cline{17-20}  
     & [X/Fe]$^{\rm a}$ & $\log\epsilon$ & $n^{\rm b}$ & s.e.  && [X/Fe]$^{\rm a}$ & $\log\epsilon$  & $n^{\rm b}$ & s.e.  && [X/Fe]$^{\rm a}$ & $\log\epsilon$ & $n^{\rm b}$ & s.e.  && [X/Fe]$^{\rm a}$ & $\log\epsilon$ & $n^{\rm b}$ & s.e.   \\
\noalign{\smallskip}
\hline
\noalign{\smallskip}
\ion{Fe}{1} & $-$2.38 & 5.12 &34 & 0.16 && $-$2.26 & 5.24 &36 & 0.11 && $-$2.55 & 4.95 &55 & 0.11 && $-$1.93 & 5.57 &58 & 0.11 \\
\ion{Fe}{2} & $-$2.38 & 5.12 & 6 & 0.20 && $-$2.25 & 5.25 & 4 & 0.16 && $-$2.55 & 4.95 & 3 & 0.15 && $-$1.93 & 5.57 &11 & 0.14 \\
\ion{Sr}{2} & $+$0.88 & 1.42 & 2 & 0.43 && $+$0.92 & 1.59 & 2 & 0.41 && $+$0.30 & 0.67 & 2 & 0.40 && $+$0.39 & 1.38 & 2 & 0.40 \\
\ion{Y}{2}  & ....    & .... & ..& .... && $+$0.73 & 0.71 & 2 & 0.20 && $+$0.59 & 0.27 & 1 & 0.20 && $+$0.16 & 0.46 & 2 & 0.18 \\
\ion{Zr}{2} & $+$1.11 & 1.34 & 1 & 0.23 && $+$1.01 & 1.37 & 2 & 0.17 && ....    & .... & ..& .... && ....    & .... & ..& .... \\
\ion{Ba}{2} & $+$2.11 & 1.95 & 2 & 0.17 && $+$2.23 & 2.20 & 3 & 0.14 && $+$1.98 & 1.65 & 1 & 0.16 && $+$1.31 & 1.60 & 1 & 0.16 \\
\ion{La}{2} & $+$1.69 & 0.53 & 6 & 0.24 && $+$2.13 & 1.10 & 5 & 0.18 && $+$2.02 & 0.69 & 3 & 0.18 && $+$1.07 & 0.36 & 3 & 0.18 \\
\ion{Ce}{2} & $+$2.01:& 1.26:& 1 & 0.22 && $+$2.13 & 1.51 &18 & 0.13 && $+$2.12 & 1.20 & 7 & 0.13 && $+$1.22 & 0.92 & 3 & 0.14 \\
\ion{Nd}{2} & $+$2.01 & 1.12 & 2 & 0.23 && $+$2.23 & 1.47 & 8 & 0.14 && $+$1.79 & 0.73 & 4 & 0.14 && $+$1.20 & 0.76 & 1 & 0.18 \\
\ion{Sm}{2} & ....    & .... & ..& .... && $+$2.08 & 0.82 & 3 & 0.15 && ....    & .... & ..& .... && ....    & .... & ..& .... \\
\ion{Eu}{2} & $+$1.73&$-$0.10& 3 & 0.20 && $+$1.88 & 0.18 & 3 & 0.15 && $+$1.62 &$-$0.38& 3 & 0.14&& $+$0.5  &$-$0.8  & 1 & 0.17 \\
\ion{Dy}{2} & ....    & .... & ..& .... && $+$1.78 & 0.70 & 2 & 0.17 && ....    & .... & ..& .... && $+$0.60&$-$0.16& 1 & 0.19 \\
\ion{Er}{2} & ....    & .... & ..& .... && $+$2.40 & 1.12 & 4 & 0.17 && ....    & .... & ..& .... && $+$1.41 & 0.45 & 1 & 0.19 \\
\ion{Pb}{2} & $+$3.3  & 3.0  & 2 & 0.24 && $+$2.84 & 2.65 & 2 & 0.19 && $+$2.4  & 1.9  & 1 & 0.19 && $+$1.9  & 2.0  & 1 & 0.19 \\
\tableline
\tableline
 &\multicolumn{4}{c}{CS~31062--050} &&\multicolumn{4}{c}{HD~196944} && \multicolumn{4}{c}{CS~22942--019} && \multicolumn{4}{c}{CS~30301--015}  \\ 
  \cline{2-5}  \cline{7-10}    \cline{12-15}   \cline{17-20}  
     & [X/Fe]$^{\rm a}$ & $\log\epsilon$ & $n^{\rm b}$ & s.e.  && [X/Fe]$^{\rm a}$ & $\log\epsilon$ & $n^{\rm b}$ & s.e.  && [X/Fe]$^{\rm a}$ & $\log\epsilon$ & $n^{\rm b}$ & s.e.  && [X/Fe]$^{\rm a}$ & $\log\epsilon$ & $n^{\rm b}$ & s.e. \\
\noalign{\smallskip}
\hline
\noalign{\smallskip}
\ion{Fe}{1} & $-$2.31 & 5.19 &26 & 0.14 && $-$2.25 & 5.25 &76 & 0.19 && $-$2.64 & 4.86 &18 & 0.14 && $-$2.64 & 4.86 &30 & 0.18 \\
\ion{Fe}{2} & $-$2.33 & 5.17 & 3 & 0.16 && $-$2.26 & 5.24 &10 & 0.20 && $-$2.64 & 4.86 & 3 & 0.15 && $-$2.63 & 4.87 & 9 & 0.15 \\
\ion{Sr}{2} & $+$0.91 & 1.51 & 2 & 0.18 && $+$0.84 & 1.51 & 3 & 0.22 && $+$1.7: & 2.0: & 2 & 0.17 && $+$0.3: & 0.6: & 1 & 0.23 \\
\ion{Y}{2}  & ....    & .... & ..& .... && $+$0.56 & 0.54 & 7 & 0.22 && $+$1.58 & 1.17 & 2 & 0.18 && $+$0.29 &$-$0.12& 4& 0.20 \\
\ion{Zr}{2} & $+$1.02 & 1.31 & 2 & 0.16 && $+$0.66 & 1.02 & 6 & 0.20 && $+$1.69 & 1.66 & 3 & 0.15 && ....    & .... & ..& .... \\
\ion{Ba}{2} & $+$2.30 & 2.21 & 2 & 0.15 && $+$1.10 & 1.07 & 3 & 0.19 && $+$1.92 & 1.50 & 3 & 0.13 && $+$1.45 & 1.03 & 1 & 0.16 \\
\ion{La}{2} & $+$2.44 & 1.34 & 4 & 0.20 && $+$0.91 &$-$0.12&6 & 0.28 && $+$1.20 &$-$0.24 & 2 & 0.26 && $+$0.84 &$-$0.58& 3& 0.25 \\
\ion{Ce}{2} & $+$2.10:& 1.41 &22 & 0.12 && $+$1.01 & 0.39 &30 & 0.19 && $+$1.54 & 0.53 & 6 & 0.12 && $+$1.16 & 0.15 &14 & 0.15 \\
\ion{Nd}{2} & $+$2.24 & 1.41 &15 & 0.12 && $+$0.86 & 0.10 &16 & 0.20 && $+$1.26 & 0.11 & 7 & 0.12 && $+$1.25 &$-$0.42& 3& 0.17 \\
\ion{Sm}{2} & $+$2.15 & 0.81 & 8 & 0.16 && $+$0.78 &$-$0.49&5 & 0.23 && $+$1.64 &$-$0.02& 1& 0.18 && $+$0.85 &$-$0.81& 3& 0.20 \\
\ion{Eu}{2} & $+$1.84 & 0.07 & 3 & 0.13 && $+$0.17 &$-$1.53&4 & 0.19 && $+$0.79 &$-$1.30& 3& 0.12 && $+$0.2: &$-$1.9:& 1& 0.18 \\
\ion{Dy}{2} & $+$2.08 & 0.93 & 1 & 0.17 && $+$0.46 &$-$0.62&1 & 0.17 && $+$0.84 &$-$0.63& 1& 0.15 && $+$0.57 &$-$0.90& 1& 0.20 \\
\ion{Er}{2} & ....    & .... & ..& .... && $+$0.81 &$-$0.47&2 & 0.22 && ....    & .... & ..& .... && ....    & .... & ..& .... \\
\ion{Pb}{2} & $+$2.9  & 2.6  & 2 & 0.24 && $+$1.9  & 1.7  & 2 & 0.24 && $\leq$1.6& $\leq$1.0&2&   && $+$1.7  & 1.1  & 2 & 0.24 \\
\tableline
\enddata
~\\

$^{\rm a}$ The value of [Fe/H] is given for Fe {\small I} and Fe {\small
II}.

$^{\rm b}$ The number of lines used for the analysis. 
\end{deluxetable}


\clearpage
\begin{figure}
\caption[]{Comparison of the observed spectra (dots) and synthetic
spectra (lines) near the Pb I 4057.8{\AA} line for HD~196944 and
CS~31062--050. Solid lines show the synthetic spectra assuming the
solar system Pb isotope ratio (see text).  The Pb abundances assumed
in the calculations are [Pb/Fe] = 1.6, 1.9, and 2.2 for HD~196944, and
[Pb/Fe] = 2.6, 2.9, and 3.2 for CS~31062--050. The dotted lines show
the synthetic spectra calculated with a single line approximation for
[Pb/Fe] = 1.9 and 3.1 for HD~196944 and CS~31062--050, respectively.}
\label{fig:sp}
\end{figure}

\begin{figure}
\caption[]{The abundances of heavy elements as a function of atomic
species for HD~196944 (upper) and CS~31062--050 (lower). The thick
solid line indicates the main {\it s}-process component determined by
\citet{arlandini99}, while the thin line indicates
the {\it r}-process component. The dotted line represents the total
solar abundance adopted in \citet{arlandini99}. All abundance patterns
are normalized to the observed Ba abundances.  }
\label{fig:abund}
\end{figure}

\begin{figure}
\caption[]{[Pb/Ba] as a function of [Fe/H]. Filled circles show the
results of the present work and of \citet{aoki02b} for LP~625--44 based
on the Subaru/HDS spectra. The open circle shows the ratio of LP~706--7
determined by \citet{norris97} and \citet{aoki01} based on the AAT/UCLES
spectrum.}
\label{fig:bapb}
\end{figure}

\clearpage
\plotone{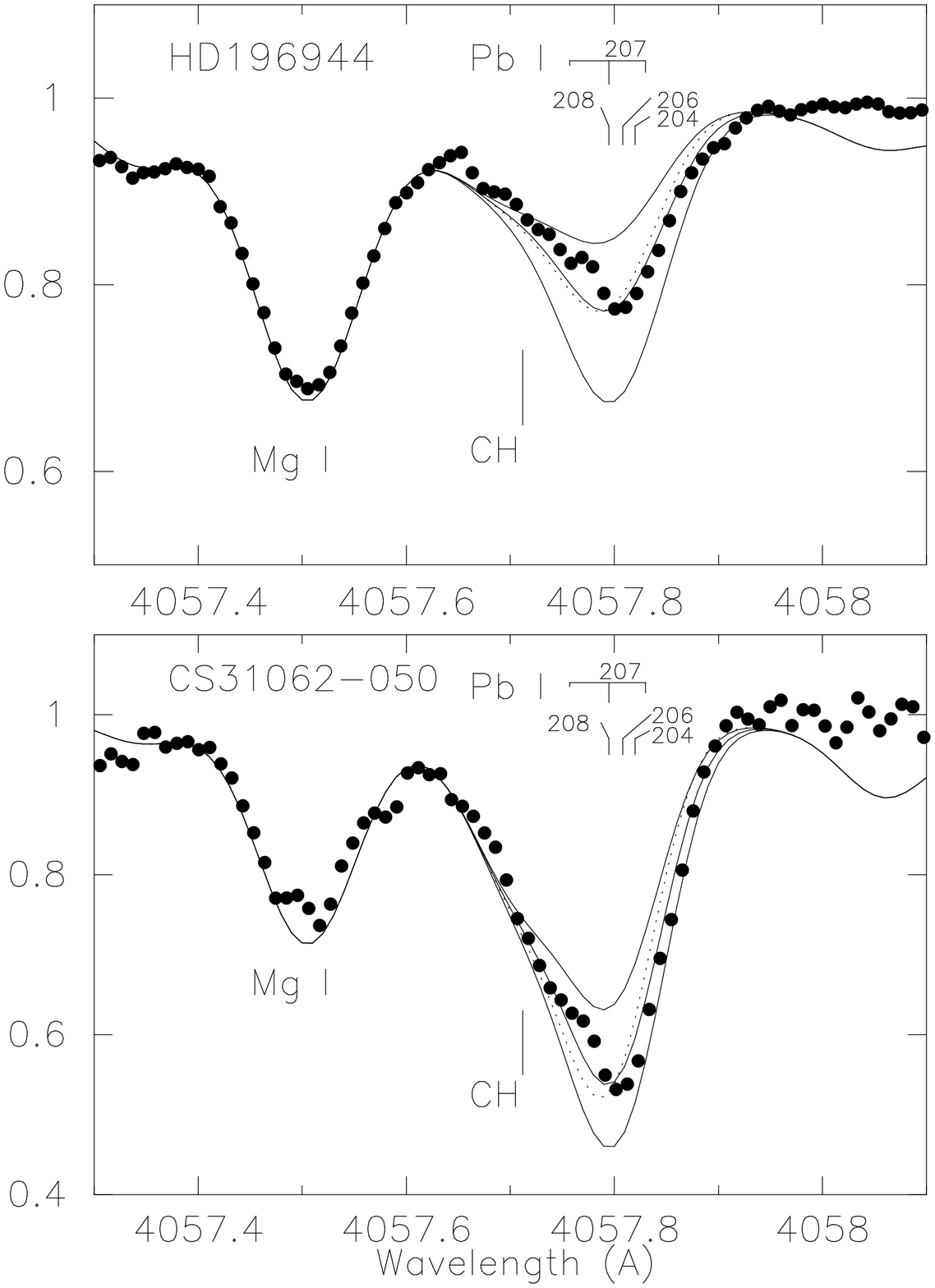}
\plotone{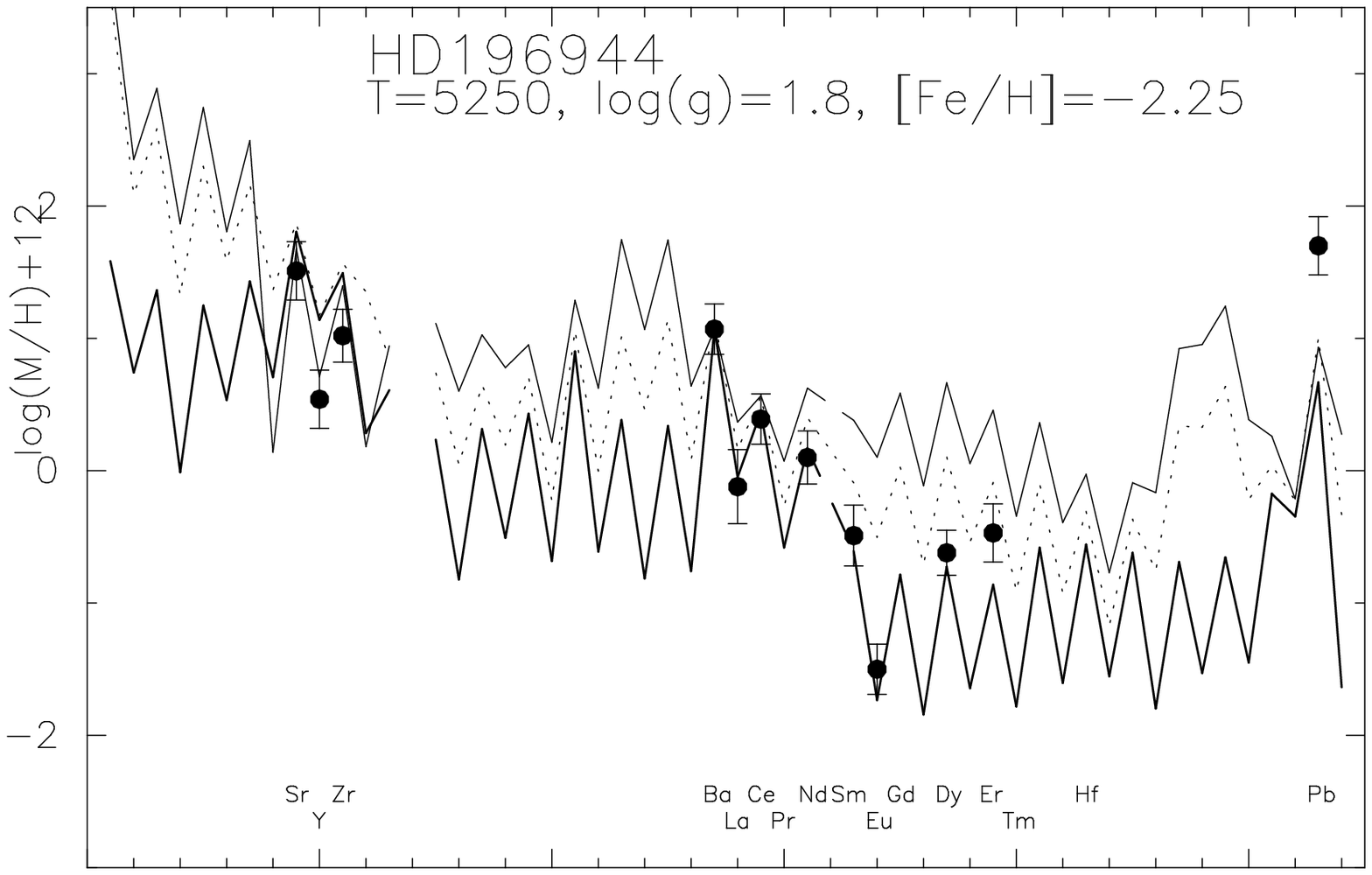}
\plotone{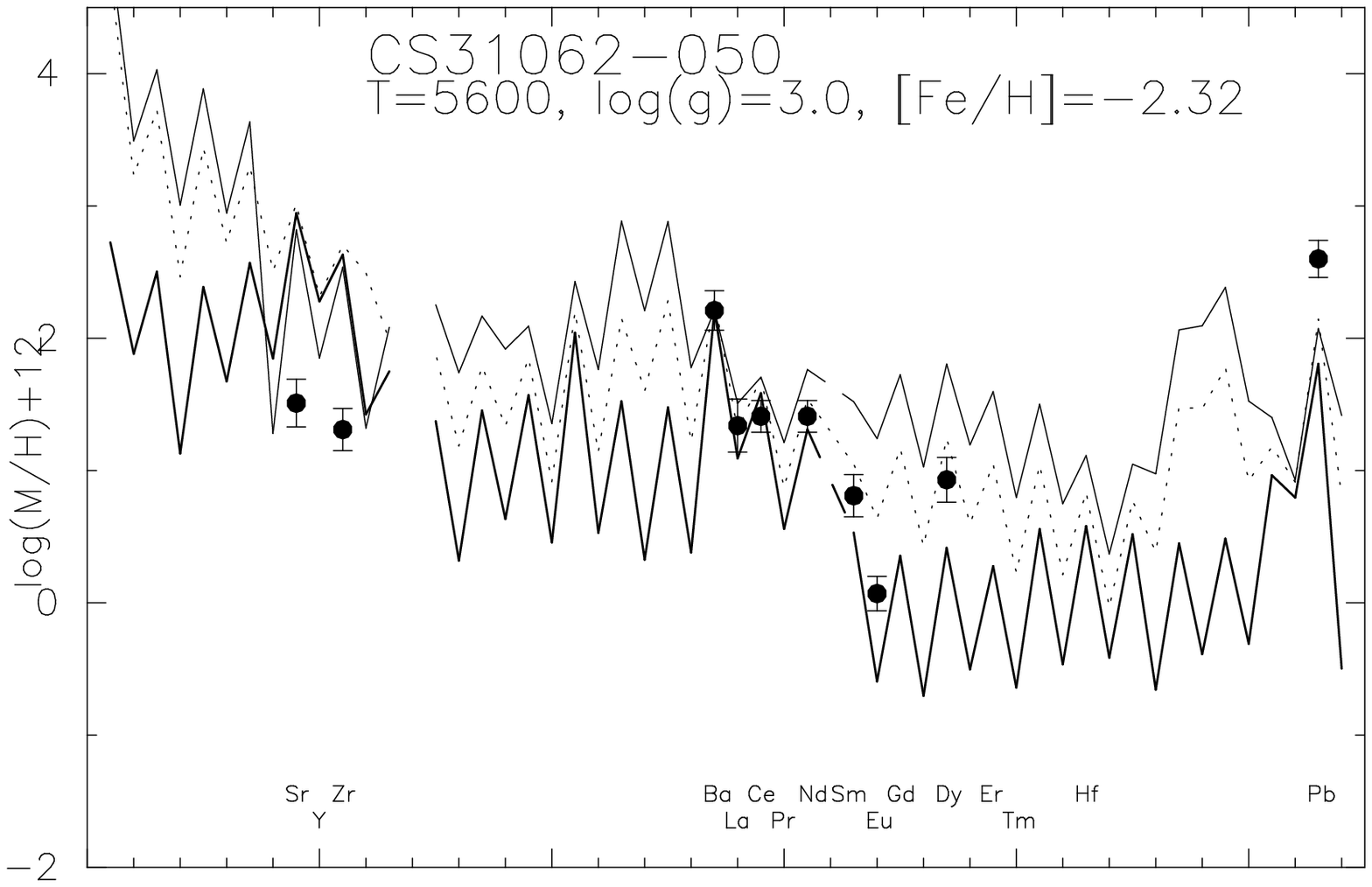}
\clearpage
\plotone{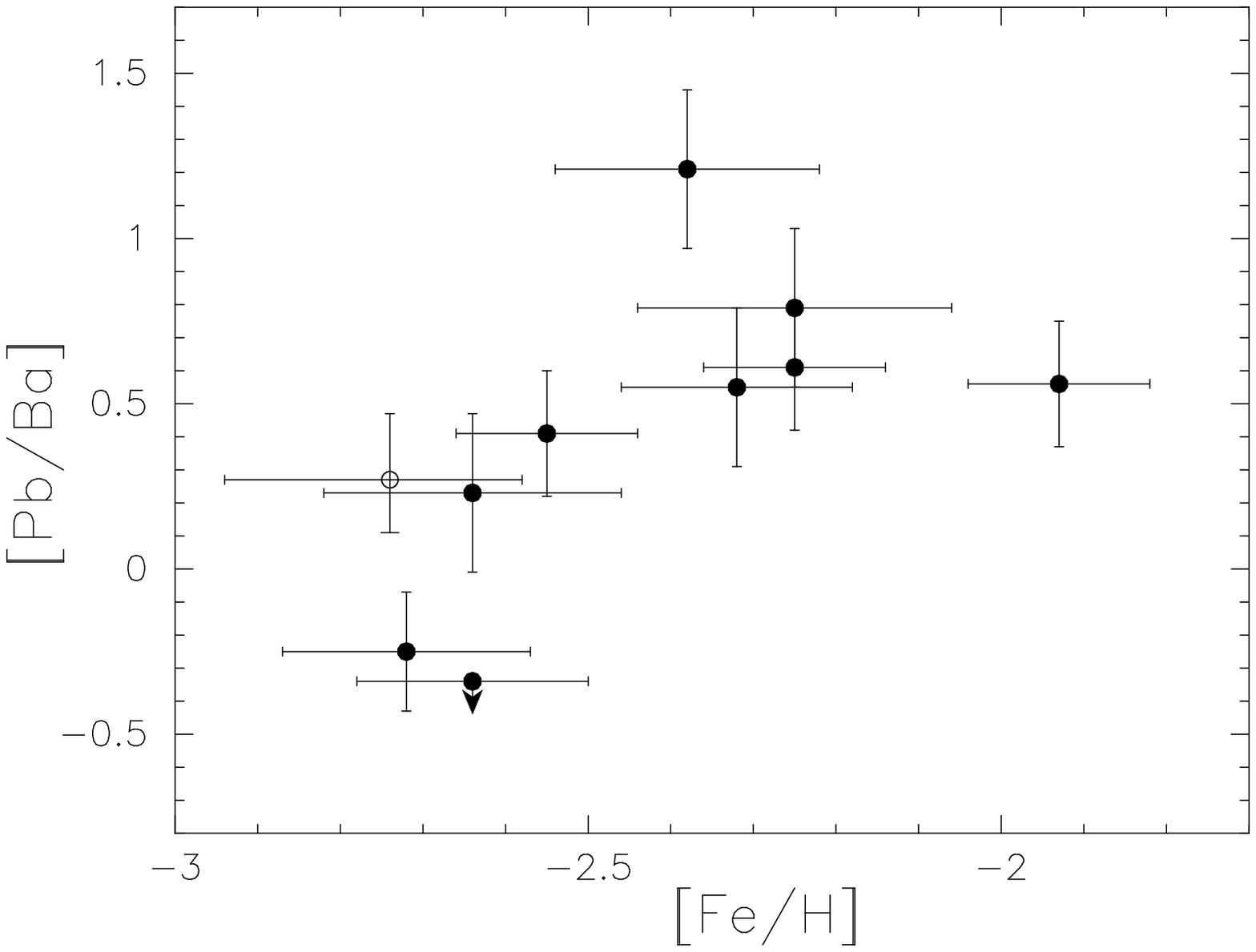}

\end{document}